# Precipitate strengthening of pyramidal slip in Mg-Zn alloys

Reza Alizadeh[1,2], Jingya Wang[1,3,4] and Javier LLorca[1,3,*]

[1]IMDEA Materials Institute, C/Eric Kandel 2, 28906 – Getafe, Madrid, Spain.
[2]Materials Science & Engineering Department, Sharif University of Technology, Tehran, Iran.
[3]Department of Materials Science, Polytechnic University of Madrid, E. T. S. de Ingenieros de Caminos, 28040 Madrid, Spain.
[4]National Engineering Research Center of Light Alloy Net Forming, Shanghai Jiao Tong University, 200240 Shanghai, PR China

*Corresponding author: javier.llorca@imdea.org

**Abstract**

The mechanical properties of Mg-4wt.% Zn alloy single crystals along the [0001] orientation were measured through micropillar compression at 23ºC and 100ºC. Basal slip was dominant in the solution treated alloy, while pyramidal slip occurred in the precipitation hardened alloy. Pyramidal dislocations pass the precipitates by forming Orowan loops, leading to homogeneous deformation and to a strong hardening. The predictions of the yield stress based on the Orowan model were in reasonable agreement with the experimental data. The presence of rod-shape precipitates perpendicular to the basal plane leads to a strong reduction in the plastic anisotropy of Mg.

## 1. Introduction

Mg and its alloys have a hexagonal close-packed (hcp) crystal structure with a *c/a* ratio close to the ideal one of 1.633. As a result, the critical resolved shear stress (CRSS) for basal slip in pure Mg is extremely low (~0.5-1 MPa) [1-2] and much lower than that required to activate non-basal slip systems. Thus, plastic deformation in Mg and its alloys is dominated by basal dislocations, and the localization of deformation in clusters of grains preferably oriented for basal slip limits the strength, ductility and formability of Mg alloys at room temperature [3]. Increasing CRSS for basal (and non-basal) slip systems is important to increase the strength of Mg alloys and to reduce the plastic anisotropy, so ductility and formability are enhanced. These objectives can be achieved by means of solution and/or precipitation hardening [4] but there are very few reliable experimental data in the literature based on mechanical tests in single crystals. Moreover, they are mainly focused on the effect of solute atoms on the CRSS for basal slip [5-8] and the information on the hardening (or softening) of pyramidal slip by solid solution is limited to Y (or Li) atoms [9-10].

These limitations have been overcome in recent years through micromechanical testing techniques of single crystal micropillars manufactured by focused ion beam milling within grains of Mg alloys with the appropriate crystallographic orientation. This approach has been successfully apply to determine CRSS of basal slip in both solution and precipitation hardened Mg-Al and Mg-Zn alloys [11-12] and also of twin nucleation and growth as well as of pyramidal slip in solution hardened Mg-Al alloys [13]. Wang and Stanford [14] carried out micropillar compression tests in single crystals of a Mg-5 wt. % Zn alloy in the as-extruded and aged conditions which were oriented to promote either basal, twinning or pyramidal slip. They did not find a large effect of precipitation on the CRSS for pyramidal slip but it should be noted that these results were obtained in micropillars of 2 μm in diameter where the size effects preclude to obtain accurate values of the CRSS [11-12]. Thus, the objective of this investigation was to determine the effect of precipitation on the CRSS for pyramidal slip in Mg-Zn alloys at room and elevated temperature (100ºC) and to ascertain the corresponding interaction mechanisms between pyramidal dislocations and precipitates.

## 2. Material and experimental techniques

Cylindrical samples of a Mg-4 wt.% Zn alloy were manufactured by casting and then subjected to a solution heat treatment at 450 °C for 20 days. Some samples were aged afterwards at 149 °C for 100 h. Micropillars of both solution treated and aged materials were milled with a focused ion beam in grains suitably oriented for pyramidal slip when deformed in compression along the [0001] direction. The micropillars had a square cross-section of 5×5 $\mu m^2$ with an aspect ratio in the range 2:1 to 3:1 [15-16]. The taper angle was always below 1º. Details of the casting process, heat treatments and micropillar milling can be found in [13].

Micropillar compression tests were carried out in a TI950 TriboindenterTM (Hysitron, Minneapolis, MN) at room temperature (23 °C) and 100 °C. The load was applied with a diamond flat punch of 10 μm in diameter. Tests were carried out under displacement control at an average strain rate of ≈$10^{-3}$ $s^{-1}$ up to a maximum strain of 10%. The experimental load-displacement curves were corrected using the Sneddon method [17] and transformed to the engineering stress-strain curves based on the initial cross-section area and height of the pillars. After deformation, the slip traces on the surface of the micropillars were analyzed in the scanning microscope. Additionally, thin lamellae of ≈100 nm in thickness were milled parallel to the micropillar axis with a focused ion beam to elucidate the deformation mechanisms by transmission Kikuchi diffraction (TKD) as well as transmission electron microscopy (TEM). More details on the mechanical characterization and lamella preparation can be found in [13].

## 3. Results and discussion

The microstructure of the alloy aged at 149ºC is shown in Fig. 1. Rod-shape precipitates have grown parallel to the *c*-axis of the Mg matrix (Fig. 1a) and their cross-section parallel to the basal plane of Mg was more or less equiaxed (Fig. 1b). Based on their morphology and orientation as well as in high-resolution TEM micrographs, they were identified as $\beta_1'$ precipitates [13], which stand for the main strengthening precipitates in Mg-Zn alloys [18-19]. The average length and diameter of the precipitates (measured from TEM images) were $l = 146 \pm 10$ nm and $d = 9.7 \pm 2$ nm, respectively. The volume fraction of the precipitates, $f$, was estimated as 2.6 ± 0.6 %. The average precipitate spacing, $\lambda$ = 82 ± 10 nm, was calculated according to [20]

$$\lambda = \sqrt{\frac{\pi}{\sqrt{3}f}}\, d \qquad (1)$$

under the hypothesis that the rods were arranged in a triangular pattern on the Mg basal plane. Zn was in solid solution in the solution treated alloy and no precipitates were found in this condition.

Micropillars were carved parallel to the [0001] axis from selected grains of both solution treated and aged samples and deformed in compression. The engineering stress-strain curves of the solution treated micropillars are plotted in Fig. 2a. The micropillar behavior was linear up to an engineering stress of >150 MPa. Afterwards, the stress-strain curves presented many load drops and reached a maximum strength of 175 - 200 MPa when the applied strain was 2%. Further deformation led to a slight but continuous reduction of the stress carried by the micropillar. This behavior was the result of the formation of many basal slip bands upon deformation, as shown in one representative scanning electron microscopy (SEM) micrograph of the deformed micropillar (Fig 2b). Thus, although the micropillars were suitably oriented to promote pyramidal slip (and the Schmid factor for pyramidal slip was 0.45), basal slip took place because the corresponding Schmid factor for basal slip was not 0 due to the unavoidable (small) misalignment between the flat punch and the micropillar upper surface. Thus, it was not possible to measure the CRSS for pyramidal slip in the solution treated Mg-Zn alloy. Nevertheless, this CRSS has to be higher than 84 ± 6 MPa because pyramidal slip did not take place when the compressive stresses were maximum in these tests.

The engineering stress-strain curves of the aged alloy deformed at 23 ºC and 100 ºC are depicted in Figs. 3a and 3b, respectively. The curves are continuous and show a linear behavior up to ≈ 400 MPa, followed by linear hardening in the plastic regime. The scatter among different samples was very small and, moreover, no significant differences are found between the samples deformed at 23 ºC and 100 ºC. SEM observation of the deformed micropillars did not show any evidence of marked slip bands on the surface (Fig. 3c). Finally, the TKD pattern obtained from a thin lamella extracted from the deformed micropillar was homogeneous and compatible with the micropillar orientation before deformation, indicating that deformation by twinning did not take place. These observations are compatible with plastic deformation due to pyramidal slip.

Assuming that <*c+a*> pyramidal slip was the dominant deformation mechanism, the yield stress could be determined from the engineering stress at the beginning of plastic deformation, where the engineering stress-strain curve deviates from the linear elastic region, following the procedure detailed in [13]. Thus, the initial CRSSs could be calculated from yield stress multiplied for the SF for pyramidal slip in the micropillar (0.45), and they can be found in Table 1. It should be noted that the CRSS for basal slip as well as for twin nucleation and growth measured in solution treated Mg-Al and Mg-Zn alloys were found to be independent of the micropillar dimensions when the cross-section of micropillars was larger than 5 x 5 µm$^2$ [11-12]. Moreover, the minimum micropillars dimensions to obtain size-independent values of the CRSS were even smaller in the case of precipitation-hardened Mg-Zn alloys oriented for basal slip [13] and Ni-based superalloys [21] because the precipitate spacing $\lambda$ was much smaller than the micropillar dimensions.

Table 1 – Initial CRSS for pyramidal slip in Mg - 4 wt. % Zn alloy aged at 149 ºC during 100 hours. The initial CRSS for basal slip for the same alloy and aging condition (from [13]) is also included for comparison

| Slip mode | Temperature (°C) | Initial CRSS (MPa) | Reference |
|---|---|---|---|
| Basal | 23 | 35.6 | [13] |
| Basal | 100 | 22.1 | [13] |
| Pyramidal | 23 | 152 ± 10 | Present work |
| Pyramidal | 100 | 155 ± 10 | Present work |

The hardening mechanisms due to the interaction of dislocations with $\beta_1'$ precipitates were studied by TEM in a thin lamella of one micropillar deformed at 23ºC. The TEM micrographs in Figs 4a and 4b with g = [0002] and g = [01$\bar{1}$1], respectively, show the presence of <*c+a*> dislocations (marked by red arrows) in the pyramidal planes whose projection is perpendicular to the precipitates in the lamella parallel to the (2$\bar{1}\bar{1}$0) plane of Mg. The STEM mode was also used to observe the dislocations at different operation vectors and the results are presented in Fig. 5, where dislocations are identified by arrows. These <*c+a*> dislocations did not shear the precipitates and seemed to make an a Orowan loop (Fig. 5a). It should be noted that evidence of precipitate shearing was not found in any of two different TEM lamellas (with approximate dimensions of 5 μm × 7 μm), which were studied almost point by point and under different beam conditions. Therefore, although clear dislocation looping could not be observed in our samples (in contrast to our previous work on this alloy for basal slip [13]), we can conclude that the <c+a> pyramidal dislocations could not shear the $\beta_1'$ precipitates and, thus, have to overcome the precipitates by the formation of Orowan loops.

It is important to notice that Jian et al. [22] reported that <c+a> pyramidal dislocations can shear the $\beta_2'$ precipitates in a ZM61 magnesium alloy (Mg-5.57wt.%Zn-0.61wt.%Mn). Thus, it should be concluded that the interaction of <c+a> dislocations with precipitates in Mg-Zn alloys depends on the orientation of the precipitates with respect to the matrix: they cannot shear the needle-like $\beta_1'$ precipitates oriented parallel to c-axis but they can shear the plate-like $\beta_2'$ precipitates oriented parallel to (0001) planes. Such different interaction mechanisms between dislocations and precipitates will affect the mechanical properties of the alloys. In particular, strong strain hardening and limited temperature dependence of the yield stress with temperature is expected when deformation is controlled by the interaction of <c+a> dislocations with $\beta_1'$ precipitates, as shown in Fig. 3. On the contrary, precipitate shearing will lead to limited strain hardening and higher influence of temperature on the yield stress, as it has been shown when deformation is controlled by the interaction of basal dislocations with $\beta_1'$ precipitates [13].

If the <c+a> dislocations surpass the precipitates through the formation of an Orowan loop, the CRSS for pyramidal slip could be calculated as $\tau_{crss} = \tau_p + \tau_{ss} + \tau_O$ where $\tau_p$ is the CRSS for pyramidal slip in pure Mg, $\tau_{ss}$ stands for the contribution from the Zn atoms in solid solution and $\tau_O$ is the Orowan contribution. The CRSS for pyramidal slip in pure Mg at room temperature is $\tau_p = 76.5$ MPa Mg [23]. Our attempts to measure the CRSS for pyramidal in the solution treated alloy were unsuccessful (see Fig. 2) and could only conclude that $\tau_p + \tau_{ss} > 84 \pm 6$ MPa in the alloy with 4 wt. % Zn in solid solution. Finally, the strengthening associated with Orowan loops around the rod-shape precipitates can be estimated as [20]

$$\tau_O = \frac{Gb}{2\pi\lambda_p\sqrt{1-\nu}} ln\frac{d_p}{r_0} \qquad (2)$$

where $G$ (=16.6 GPa) and $\nu$ (=0.33) are the shear modulus and Poisson's ratio of Mg, $b$ (= 0.33 nm) the Burgers vector, $r_0$ (=$b$) the core radius of dislocations and $d_p$ and $\lambda_p$ stand for the average diameter and spacing of the precipitates, respectively, along the $\{01\bar{1}1\}$ pyramidal plane of Mg [24]. $d_p$ (= 13.7 nm) can be easily calculated from $d$ (average diameter perpendicular to the basal plane) while $\lambda_p$ (= 114 nm) is obtained from eq. (1) using $d_p$ instead of $d$ assuming that the precipitates are distributed in a triangular pattern. Thus, the Orowan contribution to the strengthening of pyramidal slip from eq. (2) is equal to 55 MPa and $\tau_p + \tau_O$ = 131.5 MPa for pyramidal slip which is close to the experimental value measured in the micropillar compression tests at room temperature and 100ºC (Table 1). The differences between the calculated and measured values can be attributed to the solid solution strengthening due to the presence of the Zn atoms which were left in the matrix after precipitation (approximately 1.97 wt. %).

The CRSSs for basal slip at 23 ºC and 100 ºC were recently determined for the same alloy and heat treatment [13] and they were also included in Table 1. It is interesting to notice that the initial CRSS for basal slip decreased with temperature and that the stress-strain curves of compression tests in micropillars oriented for basal slip showed very little hardening after yielding and the flow stress was constant between 2% and 10% strain. TEM analysis of the deformed micropillars showed that basal dislocations could shear the precipitates, leading to the localization of deformation along one or several slip bands in the micropillar [13]. Recent molecular dynamics simulation of basal dislocation/precipitate interactions in Mg-Al [25] and Mg-Zn [26] alloys have confirmed this mechanism: basal dislocations initially surpassed the precipitates by Orowan mechanism. However, the Orowan loop penetrates the precipitate, which was finally sheared after several Orowan loops were piled-up. The number of loops necessary to shear the precipitate decreased as the temperature increased, leading to the reduction in the CRSS for basal slip with temperature, in agreement with the experimental observations in [13]. Nevertheless, precipitate shearing was not observed during the interaction of the pyramidal dislocations with $\beta'_1$ precipitates (Fig. 4). In addition, micropillars oriented to promote pyramidal slip showed a strong linear hardening after yielding, no evidence of strain localization was found and the mechanical response was independent of the temperature. All these observations are in agreement with the Orowan mechanism, which only depends on the temperature through the shear modulus $G$ (eq. 3). Finally, the predictions of the initial CRSS based on the Orowan model (eq. 2) were in good agreement with the experimental results albeit the uncertainty associated with the contribution of solid solution to the hardening of pyramidal slip.

Although the $\beta_1'$ precipitates increase more the CRSS for pyramidal slip than for basal slip, their presence leads to a strong reduction in the plastic anisotropy (understood as the ratio between the CRSS for pyramidal and basal slip) in Mg-Zn alloys, which is in the range 4.3 (23 ºC) to 7.0 (100 ºC). This is because the CRSS for basal slip in pure Mg is very low (0.5 - 1 MPa) and any mechanism that increases the magnitude will reduce the plastic anisotropy. In this respect, $\beta_1'$ precipitates increase more the CRSS for basal slip than different types of solute atoms (such as Al, Zn, or Y) [5-6,8,11] and, for instance, the plastic anisotropy of Mg-Al alloys with up to 9 wt. % of Al in solid solution is $\approx$ 13 [12], much higher than the one measured in precipitation hardened Mg-Zn alloys.

## 4. Conclusions

Compression tests were performed along the [0001] orientation in micropillars of a Mg-4 wt.% Zn alloy in the solution treated and aged conditions, the latter containing a homogeneous distribution of rod-shape precipitates parallel to the $c$ axis of the Mg lattice. Basal slip was the dominant deformation mechanisms in the solution treated alloy, while pyramidal slip occurred in the precipitation hardened alloy.

Pyramidal dislocations overcame the precipitates by the formation of Orowan loops, leading to homogeneous deformation and to a strong hardening after yielding. Moreover, the initial CRSS and the hardening rate were independent of the temperature from 23 ºC to 100 ºC, in contrast with recent results for basal slip in the same alloy, which showed evidence of precipitate shearing, localization of the deformation in slips bands after yielding and a marked reduction of the CRSS for basal slip in the same temperature range. The predictions of the initial CRSS based on the Orowan model (taking into account the contributions of the Peierls stress for pyramidal slip and of solid solution hardening) were in reasonable agreement with the experimental data.

Finally, it was shown that the presence of rod-shape precipitates perpendicular to the basal plane leads to a strong reduction in the plastic anisotropy of Mg.

**Acknowledgments**

This investigation was supported by the European Research Council (ERC) under the European Union's Horizon 2020 research and innovation programme (Advanced Grant VIRMETAL, grant agreement No. 669141. Dr. J-Y. Wang acknowledges the financial support from the China Scholarship Council (Grant no. 201506890002) and Dr. R. Alizadeh also acknowledges the support from the Spanish Ministry of Science through the Juan de la Cierva program (FJCI-2016-29660).

**Data availability**

The original data of this study can be obtained by request to the corresponding author.

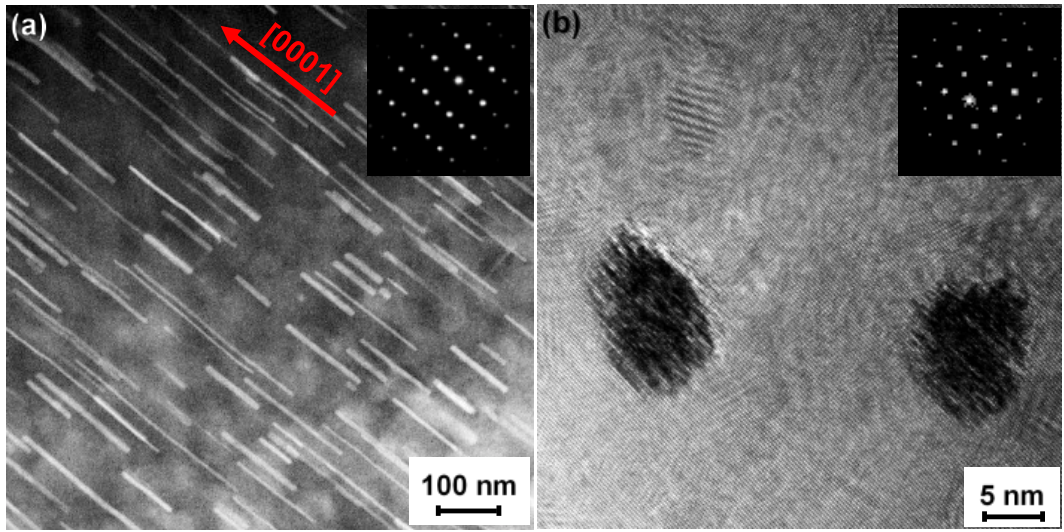

Fig. 1. (a) Scanning transmission electron micrographs of the lamella parallel to $(2\bar{1}\bar{1}0)$ plane of Mg showing the rod-shape $\beta_1'$ precipitates parallel to the *c*-axis of the matrix. (b) High resolution TEM micrograph of the lamella parallel to (0001) plane of Mg, showing the cross section of the $\beta_1'$ precipitates.

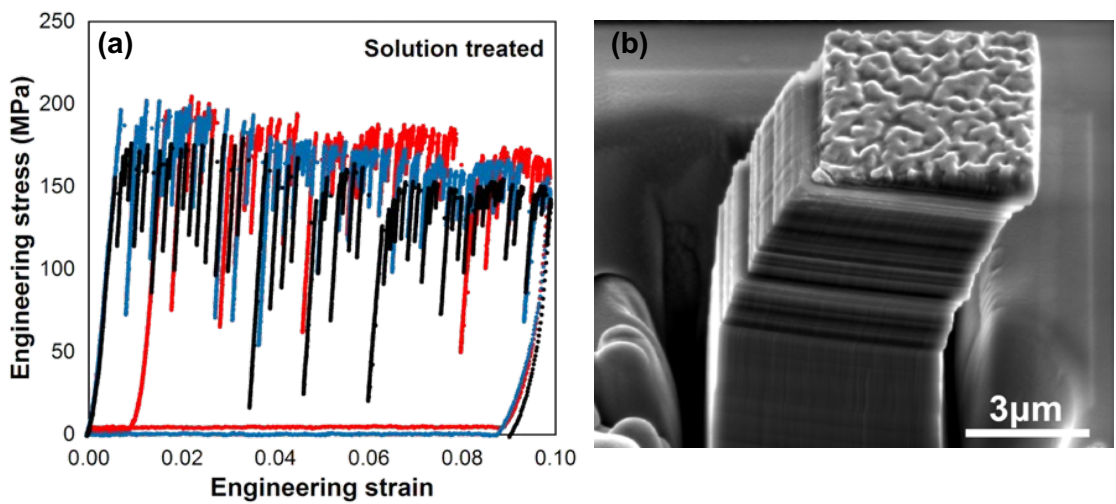

Fig. 2. (a) Engineering stress - strain curves of the three different micropillars of the solution treated alloy compressed at 23 °C along the [0001] direction. (b) SEM micrograph of one solution treated micropillar after compression at 23 °C along the [0001] direction. Deformation took place by basal slip, as shown by the slip traces.

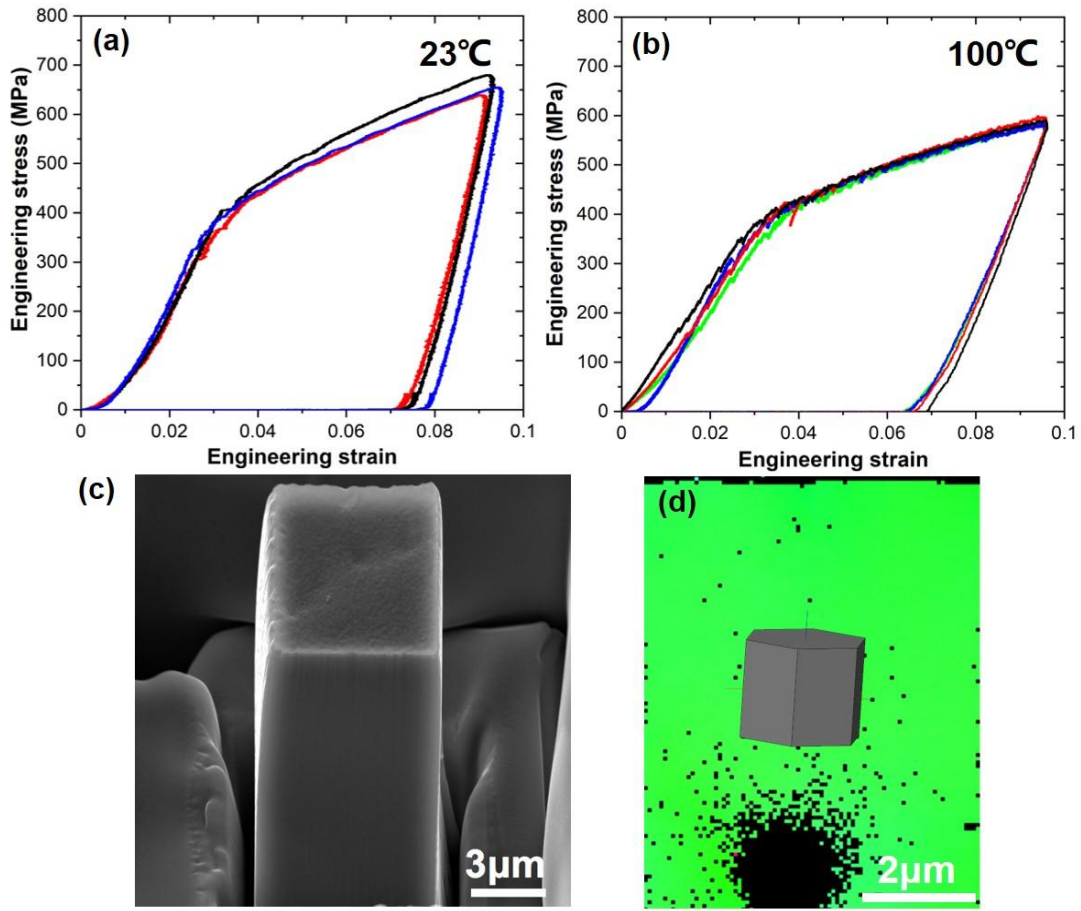

Fig. 3. (a) Engineering stress - strain curves of the micropillars of the aged alloy compressed at 23 °C along the [0001] direction. (b) Idem for micropillars deformed at 100 °C. (c) SEM micrograph of one micropillar of the aged material deformed at 23 ºC. (d) TKD pattern of one micropillar of the aged material deformed at 23 ºC indicating that twinning did not occur after compression along the [0001] direction. The black zone at the bottom of the micropillar could not be indexed due to ion beam damage during lamella extraction.

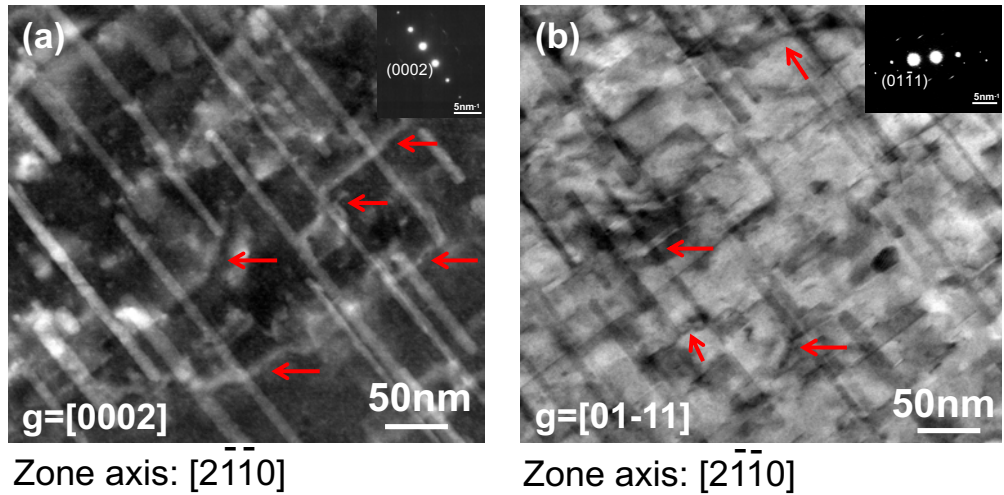

Fig. 4. Hardening mechanism ... ns with $\beta_1'$ precipitates observed by TEM in thin ... of Mg. Dark field TEM micrographs with g = [ ... ively, showing $<c+a>$ dislocations (marked w ... $\beta_1'$ precipitates. No evidence of precipitate s ... n any of the micrographs.

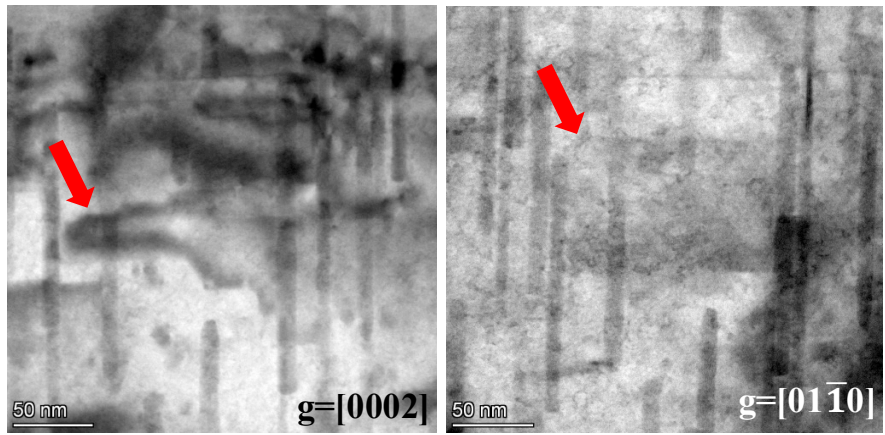

Fig. 5. STEM micrographs with g=[0002] (a) and g=[10$\bar{1}$0] (b), showing $<c+a>$ dislocations.

13